\documentclass[10pt]{article}

\usepackage{graphicx}
\usepackage{multicol}
\usepackage{psfrag}
\usepackage{subfigure}
\usepackage{amsmath,amssymb}
\usepackage{fullpage}

\newcommand{\comment}[1]{}

\title{On the Non-Termination of Ruppert's Algorithm\thanks{This note is an extension of ``On the Termination of Ruppert's Algorithm'' which appeared in the Research Notes of the 19th International Roundtable, 2010.  While generated after the submission deadline, these results were presented at the conference with the original research note.}}
\author{Alexander Rand\thanks{University of Texas-Austin, \texttt{arand@ices.utexas.edu}}}
\date{October 12, 2010 (minor revision January 3, 2011)}

\begin{document}

\maketitle

\begin{abstract}
A planar straight-line graph which causes the non-termination Ruppert's algorithm for a minimum angle threshold $\alpha \gtrapprox 29.5^\circ$ is given.  The minimum input angle of this example is about $74.5^\circ$ meaning that failure is not due to small input angles.  Additionally, a similar non-acute input is given for which Chew's second algorithm does not terminate for a minimum angle threshold $\alpha \gtrapprox 30.7^\circ$.  
\end{abstract}

For a non-acute planar straight-line graph, Ruppert's algorithm produces a conforming Delaunay triangulation composed of triangles containing no angles less than $\alpha$.  Ruppert proved the algorithm terminates for all $\alpha \lessapprox 20.7^\circ$~\cite{Ru95} and a minor addition to the analysis extends the results to input with all angles larger than $60^\circ$.  
In practice, the constraint $\alpha \lessapprox 20.7^\circ$ has been seen to be overly conservative.  
Ruppert observed that the minimum angle reaches $30^\circ$ during typical runs of the algorithm.  Further experimentation by Shewchuk \cite{Sh96} suggested that even higher values are admissible: ``In practice, the algorithm generally halts with an angle constraint of $33.8^\circ$, but often fails [at] $33.9^\circ$.'' In this note, we demonstrate an input (the upcoming Example 2) for which Ruppert's algorithm does not terminate for some minimum angle parameter $\alpha$ less than $30^\circ$.  We begin by revisiting the best known example which causes non-termination for any $\alpha > 30^\circ$.  

\begin{figure}
\centering
\psfrag{one05}{$105^\circ$}\psfrag{thirtydeg}{$30^\circ$}\psfrag{lenroot2}{$\sqrt{2}$}\psfrag{lenone}{$1$}
\includegraphics[width=.35\textwidth]{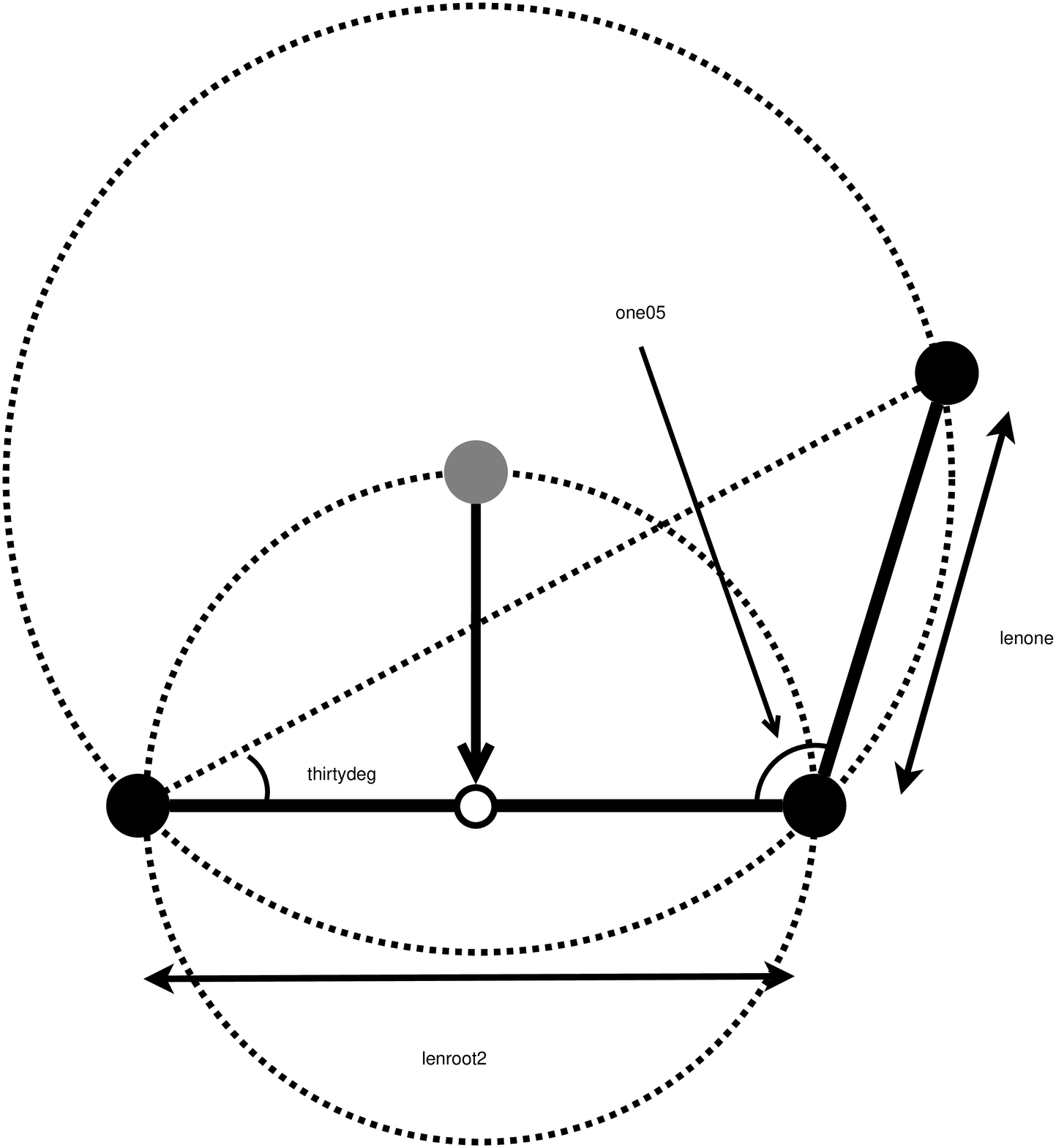}
\includegraphics[width=.35\textwidth]{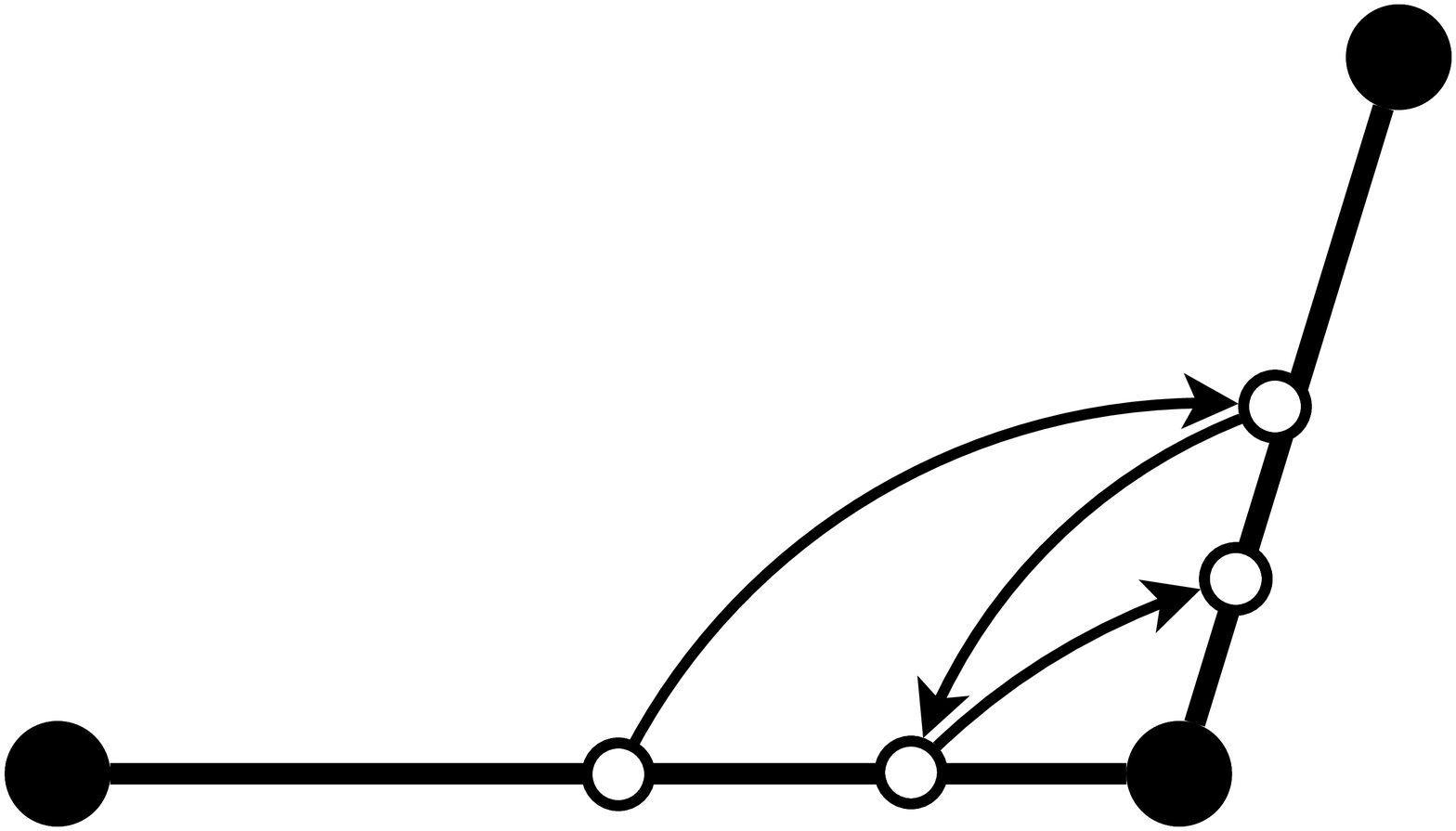}
\caption{The Pav Example demonstrates that Ruppert's algorithm may fail for any $\alpha > 30^\circ$ \cite{Pa03}.}\label{fg:pav}
\end{figure}

\paragraph{Pav Example} Steven Pav gave an example demonstrating that Ruppert's algorithm can fail to terminate for any $\alpha > 30^\circ$ \cite{Pa03}.  This example, depicted in Figure~\ref{fg:pav}, involves two adjacent segments with lengths $1$ and $\sqrt{2}$ such that they form a triangle with a $30^\circ$ angle.  Pav observed that the circumcenter of this triangle lies on the boundary of the diametral ball of the longer segment.  This causes the longer segment to split, yielding a similar configuration in which the adjacent segments are smaller by a factor of $1/\sqrt{2}$ and then repeats indefinitely.  Note: the point on the boundary of the diametral ball does not technically encroach the segment but small perturbations of this configuration yield non-termination of the algorithm for any $\alpha > 30^\circ$.

Improvements to the Pav Example face a key obstacle: each split must reduce the length of the shortest segment in the mesh by a fact of $1/\sqrt{2}$ which is the largest possible reduction according to the theory.  To eliminate this problem, we consider examples containing more adjacent input segments which allows for a smaller reduction in segment length at each midpoint insertion.  However, throughout these examples we maintain our requirement that input angles are larger than $60^\circ$.  

\begin{figure}
\centering
\psfrag{theangle}{$30.7^\circ$}\psfrag{len234}{$2$}\psfrag{len1}{$2^{1/4}$}
\includegraphics[width=.35\textwidth]{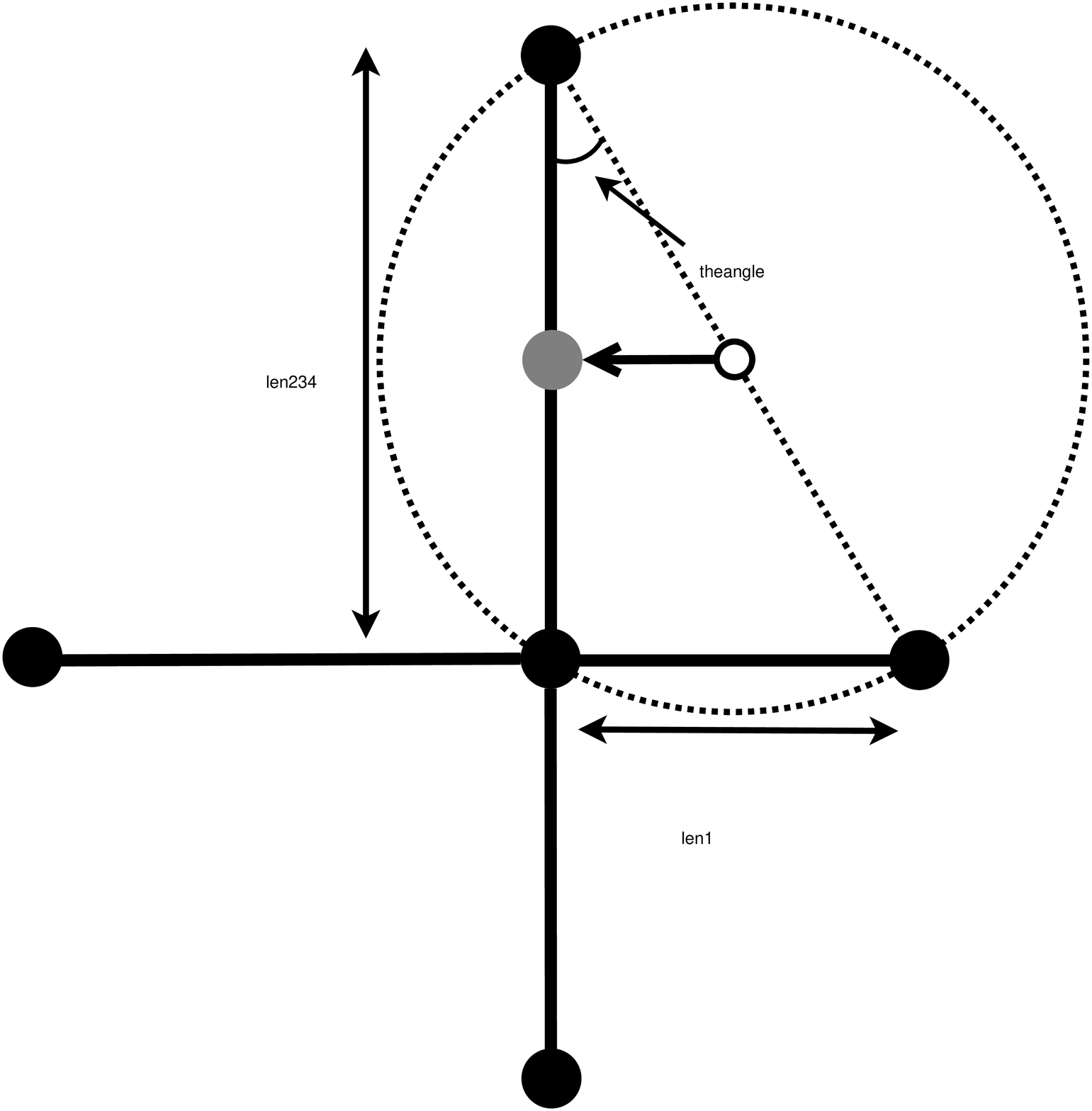}
\psfrag{j1}{}\psfrag{j2}{}\psfrag{j3}{}\psfrag{j4}{}\hspace{.1in}
\includegraphics[width=.35\textwidth]{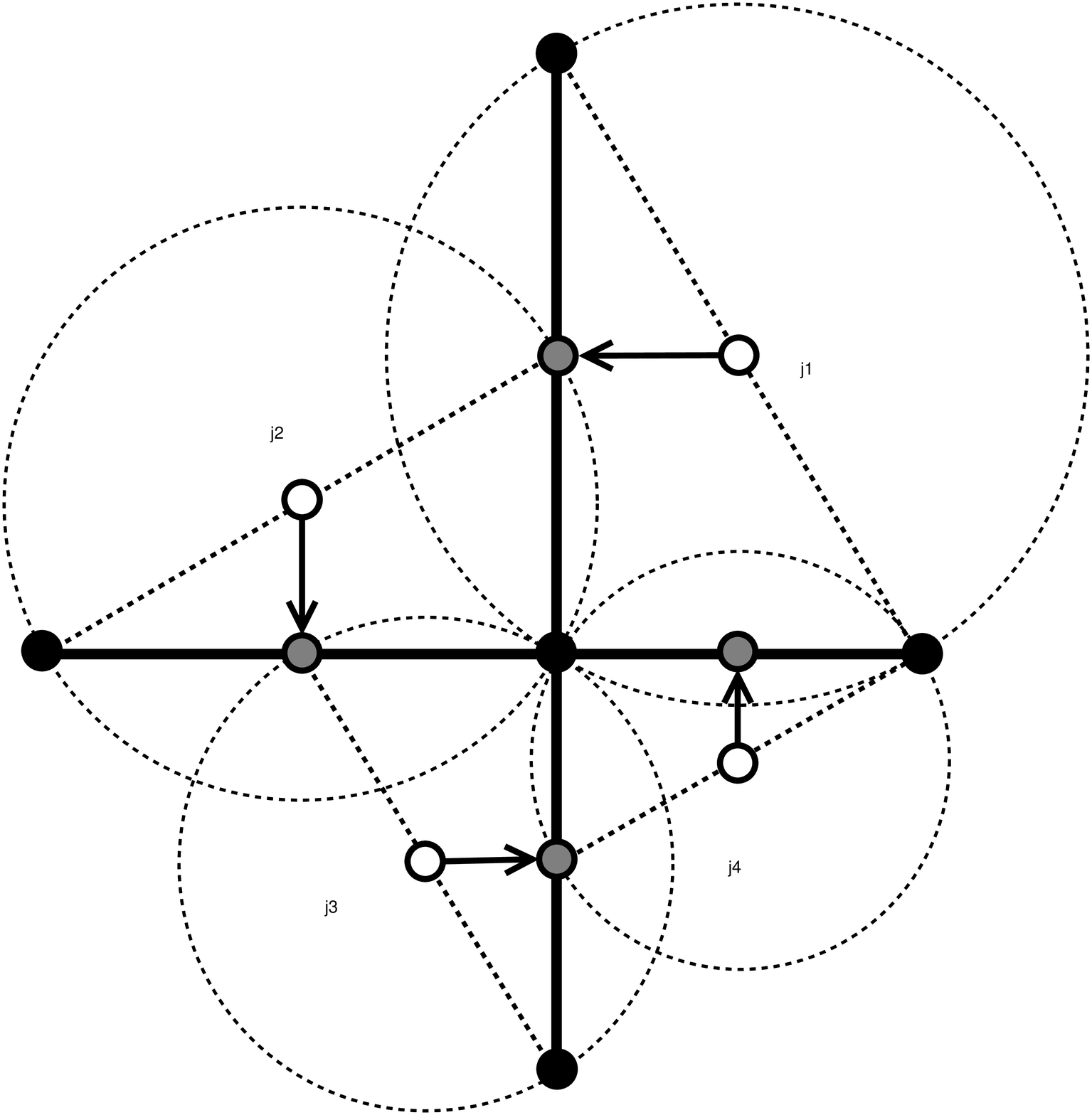}
\caption{Example 1 demonstrating non-termination of Ruppert's algorithm (and Chew's second algorithm) for ${\alpha \gtrapprox 30.7^\circ}$.}\label{fg:pin4}
\end{figure}

\paragraph{Example 1} Consider a non-acute input containing four adjacent segments of lengths $2$, $2^{3/4}$, $2^{1/2}$ and $2^{1/4}$ as in Fig.~\ref{fg:pin4}.  
The endpoints of the longest and shortest segments form a Delaunay triangle with smallest angle $\arctan 2^{-3/4}\approx 30.7^\circ$.  If $\alpha > \arctan 2^{-3/4}$, the circumcenter of this triangle encroaches upon the longer segment causing the midpoint of the longest segment to be inserted.  Now the adjacent segments have lengths $1$, $2^{3/4}$, $2^{1/2}$ and $2^{1/4}$ and the ratio of the shortest and longest segment is still $2^{3/4}$.  Again this gives a poor quality triangle and the midpoint of the longest segment is inserted.  This cycle repeats indefinitely. 

While the Pav Example provides a sharper limit on the performance of Ruppert's algorithm, Example 1 has importance of its own.  Chew's second algorithm~\cite{Ch93} (widely used as the default mesh generator in Triangle~\cite{Sh02}) terminates on the Pav Example while it fails to terminate on Example 1.  

\begin{figure}
\centering
\psfrag{len213}{$2^{1/3}$}\psfrag{lentwo}{$2$}\psfrag{pin3ang}{$22^\circ$}
\includegraphics[width=.42\textwidth]{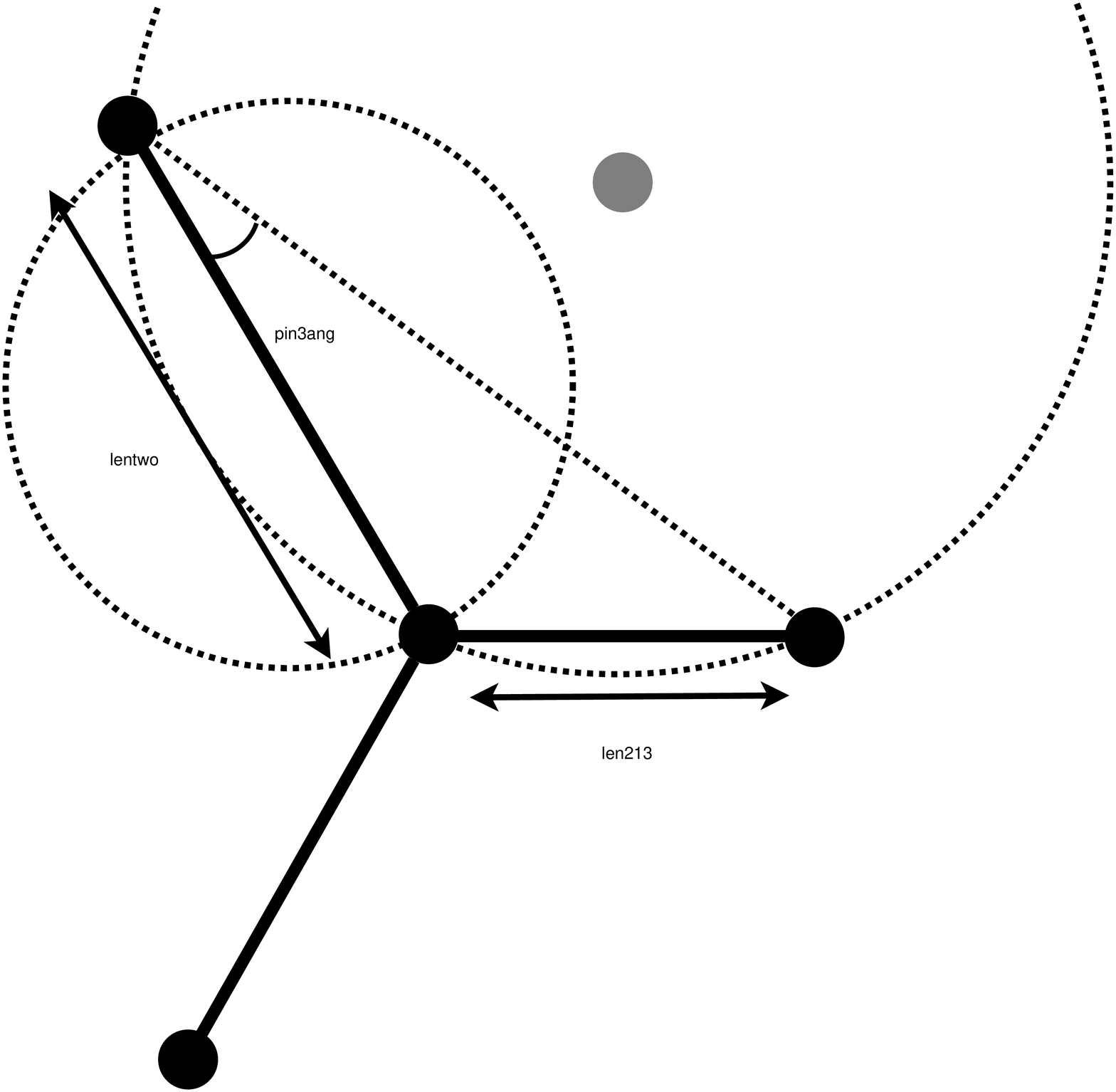}
\hspace{.02\textwidth}
\psfrag{len215}{$2^{1/5}$}\psfrag{lentwo}{$2$}\psfrag{pin5ang}{$33^\circ$}
\includegraphics[width=.33\textwidth]{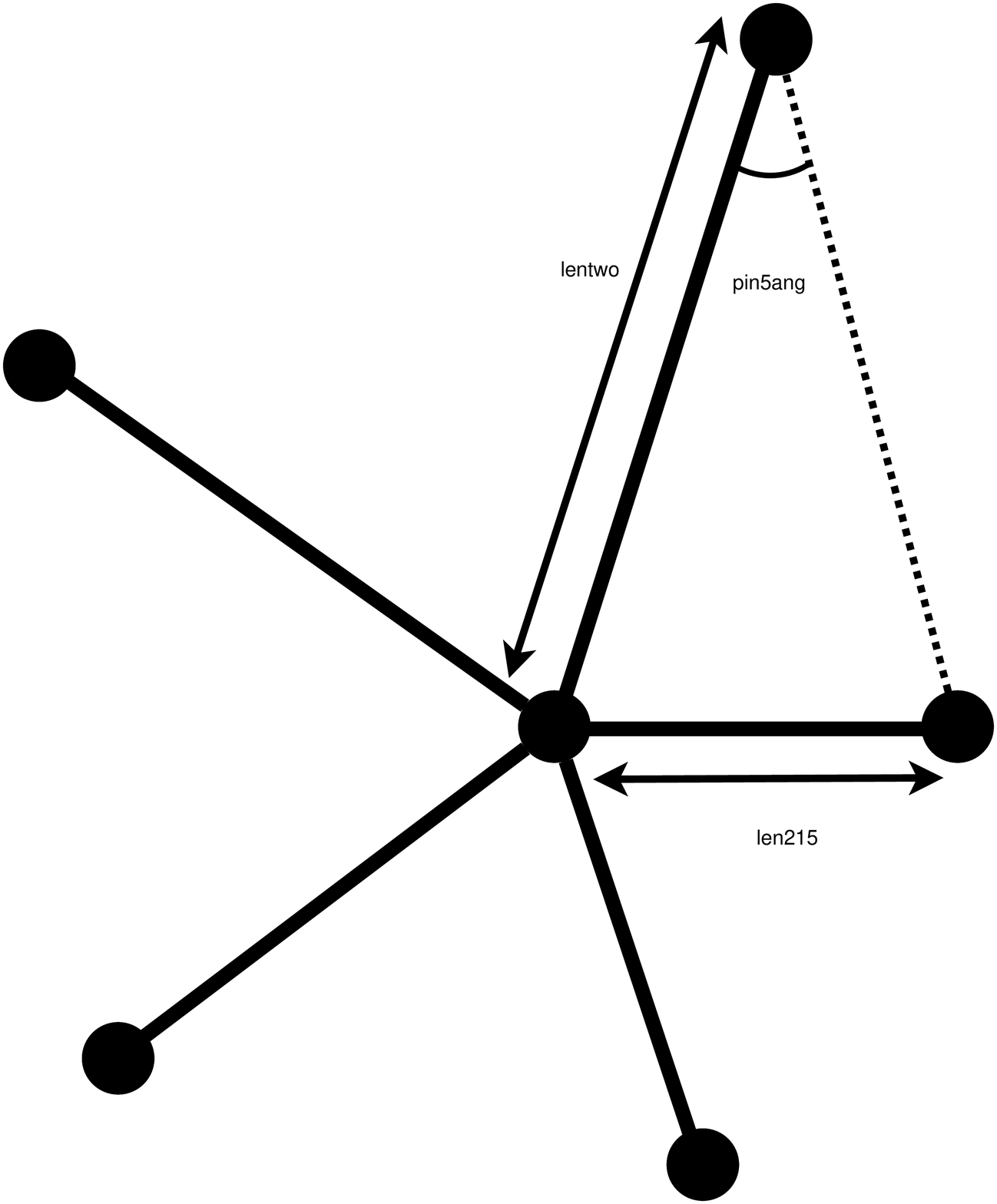}
\caption{Example input with three and five adjacent end segments.  Left: The three segment version does not lead to indefinite refinement. Right: the five segment variant leads to a weaker restriction on the minimum angle threshold.}\label{fg:pin35}
\end{figure}

Following the approach of Example 1, input with three and five adjacent segments can be constructed seeking non-termination; see Figure~\ref{fg:pin35}.  In the three segment version the skinny triangle circumcenter does not encroach the longest segment and no cascading encroachment occurs.  The five segment variant does lead to non-termination but only for an angle threshold larger than about $33^\circ$ and thus does not improve upon the previous results.  

\begin{figure}
\centering
\psfrag{angle30}{$30^\circ$}\psfrag{lenroot2}{$\sqrt{2}$}\psfrag{angle29}{$29^\circ$}\psfrag{lentwo}{$2$}\psfrag{angle75}{$75^\circ$}\psfrag{len1}{$1$}
\includegraphics[width=.45\textwidth]{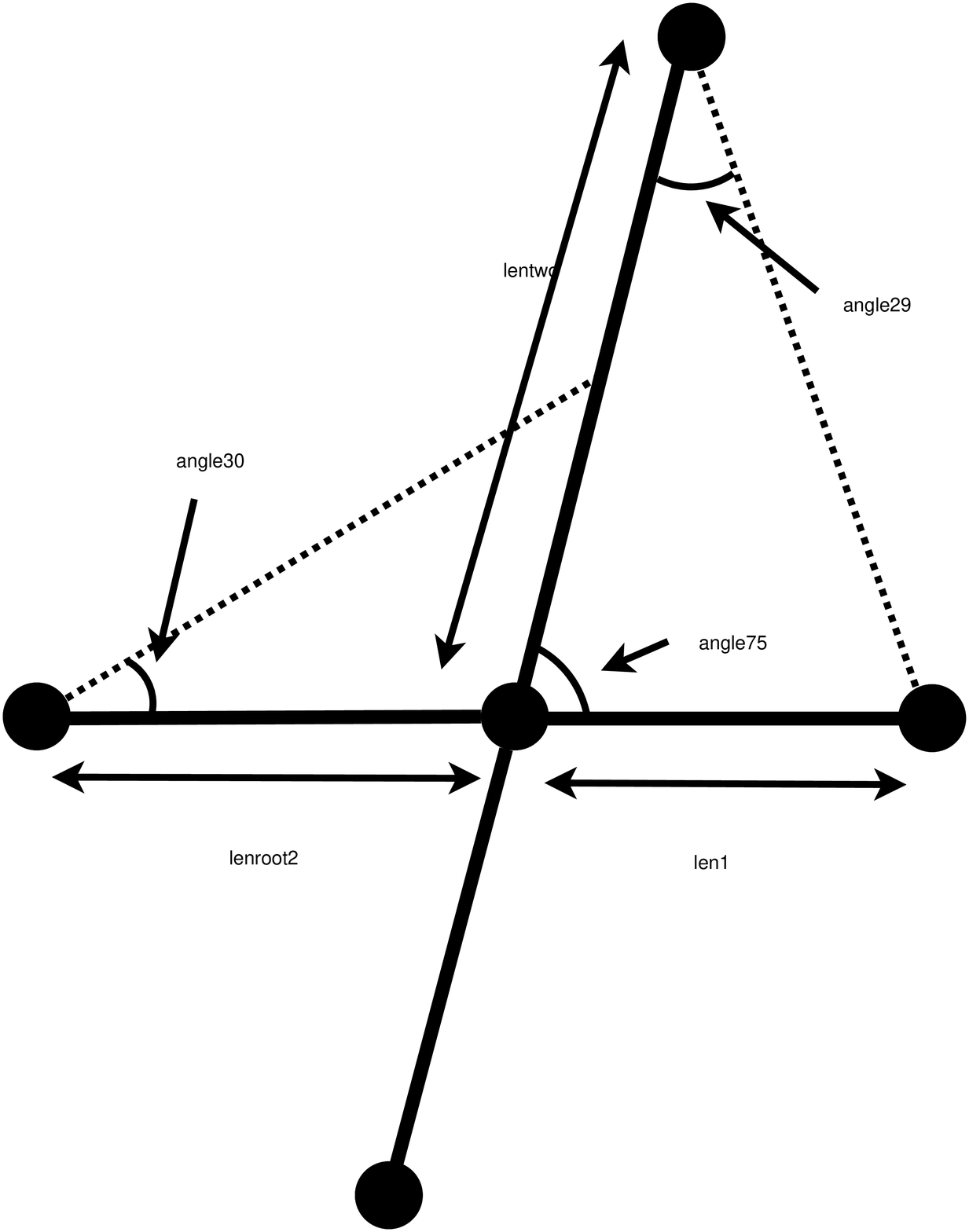}
\psfrag{alph1}{$\alpha_1$}\psfrag{alph2}{$\alpha_2$}\psfrag{lena}{$2a$}\psfrag{theta}{$\theta$}
\includegraphics[width=.45\textwidth]{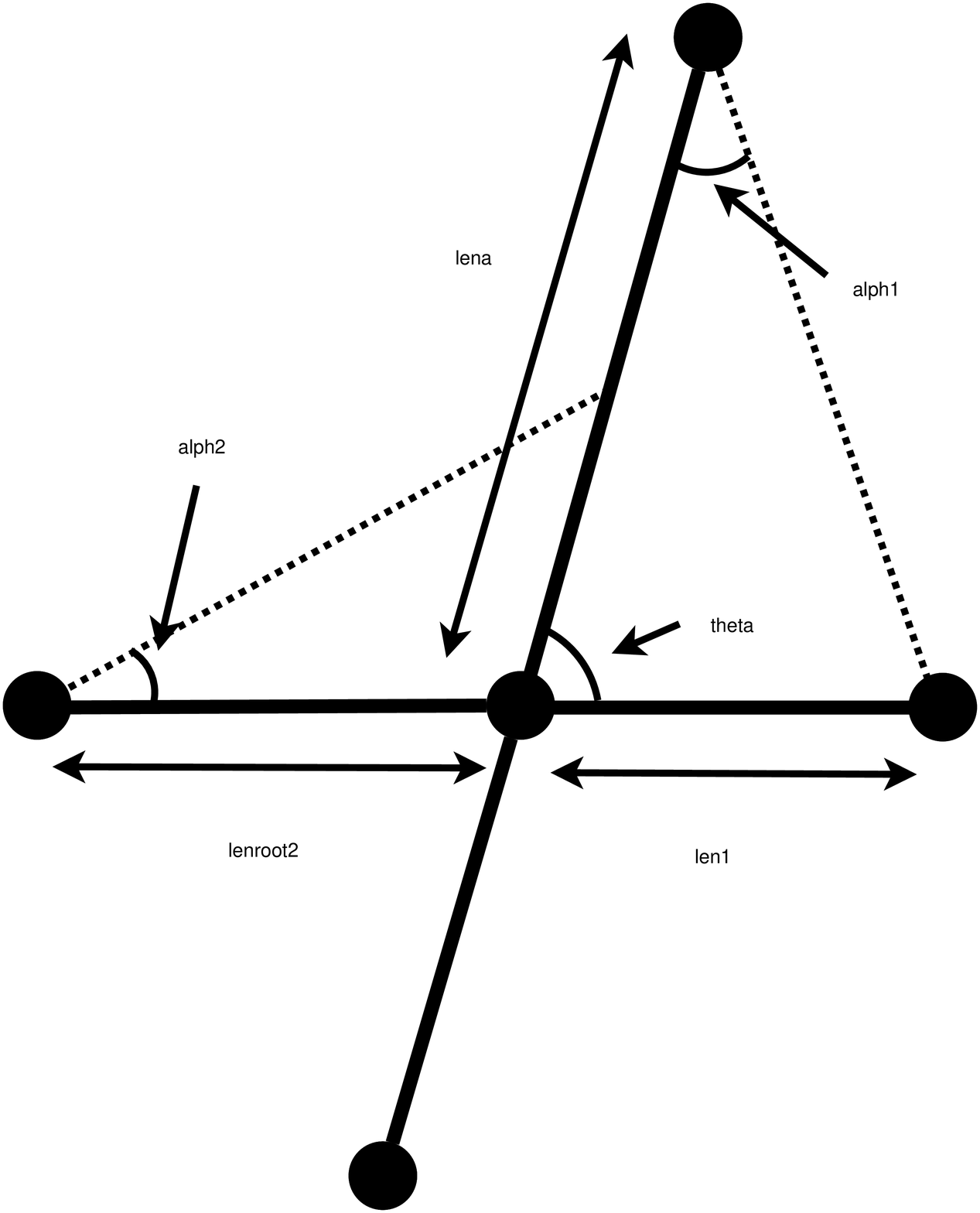}
\caption{Left: a combination of the input in the Pav Example and Example 1.  In this configuration, Ruppert's algorithm still fails to terminate for $\alpha > 30^\circ$.  Right: a generalization of the example to the left.  $\theta$ and $a$ can be varied to minimize the maximum of $\alpha_1$ and $\alpha_2$ while requiring that an infinite encroachment sequence occurs.}\label{fg:combo}
\end{figure}

\paragraph{Example 2}

A better example can be generated by combining ideas in the Pav Example and Example 1.  First consider a four segment input that contains two adjacent copies of Example 1 as shown in Figure~\ref{fg:combo}(left).  Thanks to a fortuitous angle of about $29^\circ$ (at the top of the figure) non-termination can occur for any $\alpha > 30^\circ$ by spiraling around all four input segments rather than just alternating between two segments as in the Pav Example.  

\begin{figure}
\centering
\psfrag{angle30}{$30^\circ$}\psfrag{lenroot2}{$\sqrt{2}$}\psfrag{angle29}{$29^\circ$}\psfrag{lentwo}{$2$}\psfrag{angle75}{$75^\circ$}\psfrag{len1}{$1$}
\psfrag{angle29a}{$29.5^\circ$}\psfrag{lentwoa}{$1.97$}\psfrag{angle75a}{$74.5^\circ$}
\includegraphics[width=.45\textwidth]{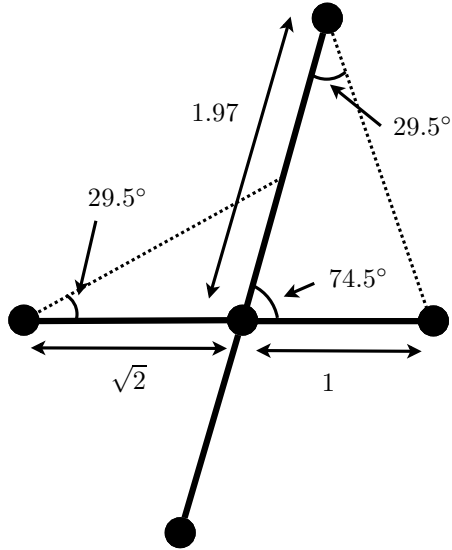}
\caption{Best known example for which Ruppert's algorithm fails to terminate without small input angles.}\label{fg:opt}
\end{figure}

Finally, improvement is made by perturbing this example to balance the $30^\circ$ and $29^\circ$ angles.  As shown in Figure~\ref{fg:combo}(right) the length of  the upper segment ($2a$) as well as the angle between two segments ($\theta$) are allowed to vary while still requiring that the desired encroachment occurs.  Our goal is to minimize the larger of the small angles in the skinny triangles ($\alpha_1$ and $\alpha_2$ in Figure~\ref{fg:combo}(right)).  The minimum value of $\max(\alpha_1,\alpha_2)$ occurs at a solution of the following four equations.
\begin{align*}
\sin \theta & = \cos\theta + a/\sqrt{2}\\
\cos \theta & = 2a - \cos\alpha_1 \sqrt{4a^2 + 1 - 4a\cos\theta}\\
\sin \theta & = \tan \alpha_2 (\cos \theta + \sqrt{2}/a)\\
\alpha_1 & = \alpha_2
\end{align*}
Searching near the original values $\theta = 75^\circ$, $a = 1$, $\alpha_1 = 29^\circ$, and $\alpha_2 = 30^\circ$ gives a numerical solution:
\begin{align*}
\theta & = 74.51^\circ, & \alpha_{1,2} &= 29.51^\circ, & {\rm and}  & & a &= 0.985.
\end{align*}
Thus for this configuration, depicted in Figure~\ref{fg:opt}, Ruppert's algorithm fails to terminate for any $\alpha \gtrapprox 29.51^\circ$.

%
% Solution from matlab
%
%   theta = 74.505444359244166   
%   alpha = 29.505445544437830    
%       a = 0.985012525482730

\bibliographystyle{abbrv}
\bibliography{ruppert}

\end{document}